\magnification=\magstep1


\newbox\SlashedBox
\def\slashed#1{\setbox\SlashedBox=\hbox{#1}
\hbox to 0pt{\hbox to 1\wd\SlashedBox{\hfil/\hfil}\hss}{#1}}
\def\hboxtosizeof#1#2{\setbox\SlashedBox=\hbox{#1}
\hbox to 1\wd\SlashedBox{#2}}

\def\mathslashed#1{\setbox\SlashedBox=\hbox{$#1$}
\hbox to 0pt{\hbox to 1\wd\SlashedBox{\hfil/\hfil}\hss}#1}

\def\ifsmall{\iffalse}  
\def\titlepagefont{}  

\def\DefineTeXgraphics{%
\special{ps::[global] /TeXgraphics { } def}}  

\def\today{\ifcase\month\or January\or February\or March\or April\or May
\or June\or July\or August\or September\or October\or November\or
December\fi\space\number\day, \number\year}
\def\eatPrefix19{}
\def\Year{\expandafter\eatPrefix\the\year}
\newcount\hours \newcount\minutes
\def\monthname{\ifcase\month\or
January\or February\or March\or April\or May\or June\or July\or
August\or September\or October\or November\or December\fi}
\def\shortmonthname{\ifcase\month\or
Jan\or Feb\or Mar\or Apr\or May\or Jun\or Jul\or
Aug\or Sep\or Oct\or Nov\or Dec\fi}

\def\TimeStamp{\hours\the\time\divide\hours by60%
\minutes -\the\time\divide\minutes by60\multiply\minutes by60%
\advance\minutes by\the\time%
${\rm \shortmonthname}\cdot\if\day<10{}0\fi\the\day\cdot\the\year%
\qquad\the\hours:\if\minutes<10{}0\fi\the\minutes$}







\newif\ifdraftmode
\newif\ifleftlabels  

\def\nolabels{\def\wrlabeL##1{}\def\eqlabeL##1{}\def\reflabeL##1{}}
\def\writelabels{\def\wrlabeL##1{\leavevmode\vadjust{\rlap{\smash%
{\line{{\escapechar=` \hfill\rlap{\sevenrm\hskip.03in\string##1}}}}}}}%
\def\eqlabeL##1{{\escapechar-1\rlap{\sevenrm\hskip.05in\string##1}}}%
\def\reflabeL##1{\noexpand\rlap{\noexpand\sevenrm[\string##1]}}}
\def\writeleftlabels{\def\wrlabeL##1{\leavevmode\vadjust{\rlap{\smash%
{\line{{\escapechar=` \hfill\rlap{\sevenrm\hskip.03in\string##1}}}}}}}%
\def\eqlabeL##1{{\escapechar-1%
\rlap{\sixrm\hskip.05in\string##1}%
\llap{\sevenrm\string##1\hskip.03in\hbox to \hsize{}}}}%
\def\reflabeL##1{\noexpand\rlap{\noexpand\sevenrm[\string##1]}}}
\nolabels

\newdimen\fullhsize
\newdimen\hstitle
\hstitle=\hsize 
\newdimen\hsbody
\hsbody=\hsize 
\newdimen\hbodyoffset
\hbodyoffset=\hoffset 
\newbox\leftpage
\def\abstract#1{#1}
\def\rotated{\special{ps: landscape}
\magnification=1200  
\baselineskip=14pt
\global\hstitle=9truein\global\hsbody=4.75truein
\global\vsize=7truein\global\voffset=-.31truein
\global\hoffset=-0.54in\global\hbodyoffset=-.54truein
\global\fullhsize=10truein
\def\DefineTeXgraphics{%
\special{ps::[global]
/TeXgraphics {currentpoint translate 0.7 0.7 scale
              -80 0.72 mul -1000 0.72 mul translate} def}}
\let\lr=L
\def\ifsmall{\iftrue}
\def\titlepagefont{\twelvepoint}
\trueseventeenpoint
\def\almostshipout##1{\if L\lr \count1=1
      \global\setbox\leftpage=##1 \global\let\lr=R
   \else \count1=2
      \shipout\vbox{\hbox to\fullhsize{\box\leftpage\hfil##1}}
      \global\let\lr=L\fi}

\output={\ifnum\count0=1 
 \shipout\vbox{\hbox to \fullhsize{\hfill\pagebody\hfill}}\advancepageno
 \else
 \almostshipout{\leftline{\vbox{\pagebody\makefootline}}}\advancepageno
 \fi}

\def\abstract##1{{\leftskip=1.5in\rightskip=1.5in ##1\par}} }

\def\linemessage#1{\immediate\write16{#1}}

\global\newcount\secno \global\secno=0
\global\newcount\appno \global\appno=0
\global\newcount\meqno \global\meqno=1
\global\newcount\subsecno \global\subsecno=0
\global\newcount\figno \global\figno=0

\newif\ifAnyCounterChanged
\let\terminator=\relax
\def\normalize#1{\ifx#1\terminator\let\next=\relax\else%
\if#1i\aftergroup i\else\if#1v\aftergroup v\else\if#1x\aftergroup x%
\else\if#1l\aftergroup l\else\if#1c\aftergroup c\else%
\if#1m\aftergroup m\else%
\if#1I\aftergroup I\else\if#1V\aftergroup V\else\if#1X\aftergroup X%
\else\if#1L\aftergroup L\else\if#1C\aftergroup C\else%
\if#1M\aftergroup M\else\aftergroup#1\fi\fi\fi\fi\fi\fi\fi\fi\fi\fi\fi\fi%
\let\next=\normalize\fi%
\next}
\def\makeNormal#1#2{\def\doNormalDef{\edef#1}\begingroup%
\aftergroup\doNormalDef\aftergroup{\normalize#2\terminator\aftergroup}%
\endgroup}

\def\warnIfChanged#1#2{%
\ifundef#1
\else\begingroup%
\edef\oldDefinitionOfCounter{#1}\edef\newDefinitionOfCounter{#2}%
\ifx\oldDefinitionOfCounter\newDefinitionOfCounter%
\else%
\linemessage{Warning: definition of \noexpand#1 has changed.}%
\global\AnyCounterChangedtrue\fi\endgroup\fi}

\def\Section#1{\global\advance\secno by1\relax\global\meqno=1%
\global\subsecno=0%
\bigbreak\bigskip
\centerline{\twelvepoint \bf %
\the\secno. #1}%
\par\nobreak\medskip\nobreak}
\def\tagsection#1{%
\warnIfChanged#1{\the\secno}%
\xdef#1{\the\secno}%
\ifWritingAuxFile\immediate\write\auxfile{\noexpand\xdef\noexpand#1{#1}}\fi%
}
\def\section{\Section}
\def\Subsection#1{\global\advance\subsecno by1\relax\medskip %
\leftline{\bf\the\secno.\the\subsecno\ #1}%
\par\nobreak\smallskip\nobreak}
\def\tagsubsection#1{%
\warnIfChanged#1{\the\secno.\the\subsecno}%
\xdef#1{\the\secno.\the\subsecno}%
\ifWritingAuxFile\immediate\write\auxfile{\noexpand\xdef\noexpand#1{#1}}\fi%
}

\def\subsection{\Subsection}

\def\romappno{\uppercase\expandafter{\romannumeral\appno}}
\def\makeNormalizedRomappno{%
\expandafter\makeNormal\expandafter\normalizedromappno%
\expandafter{\romannumeral\appno}%
\edef\normalizedromappno{\uppercase{\normalizedromappno}}}
\def\Appendix#1{\global\advance\appno by1\relax\global\meqno=1\global\secno=0%
\global\subsecno=0%
\bigbreak\bigskip
\centerline{\twelvepoint \bf Appendix %
\romappno. #1}%
\par\nobreak\medskip\nobreak}
\def\tagappendix#1{\makeNormalizedRomappno%
\warnIfChanged#1{\normalizedromappno}%
\xdef#1{\normalizedromappno}%
\ifWritingAuxFile\immediate\write\auxfile{\noexpand\xdef\noexpand#1{#1}}\fi%
}
\def\appendix{\Appendix}
\def\Subappendix#1{\global\advance\subsecno by1\relax\medskip %
\leftline{\bf\romappno.\the\subsecno\ #1}%
\par\nobreak\smallskip\nobreak}
\def\tagsubappendix#1{\makeNormalizedRomappno%
\warnIfChanged#1{\normalizedromappno.\the\subsecno}%
\xdef#1{\normalizedromappno.\the\subsecno}%
\ifWritingAuxFile\immediate\write\auxfile{\noexpand\xdef\noexpand#1{#1}}\fi%
}

\def\eqn#1{\makeNormalizedRomappno%
\ifnum\secno>0%
  \warnIfChanged#1{\the\secno.\the\meqno}%
  \eqno(\the\secno.\the\meqno)\xdef#1{\the\secno.\the\meqno}%
     \global\advance\meqno by1
\else\ifnum\appno>0%
  \warnIfChanged#1{\normalizedromappno.\the\meqno}%
  \eqno({\rm\romappno}.\the\meqno)%
      \xdef#1{\normalizedromappno.\the\meqno}%
     \global\advance\meqno by1
\else%
  \warnIfChanged#1{\the\meqno}%
  \eqno(\the\meqno)\xdef#1{\the\meqno}%
     \global\advance\meqno by1
\fi\fi%
\eqlabeL#1%
\ifWritingAuxFile\immediate\write\auxfile{\noexpand\xdef\noexpand#1{#1}}\fi%
}
\def\defeqn#1{\makeNormalizedRomappno%
\ifnum\secno>0%
  \warnIfChanged#1{\the\secno.\the\meqno}%
  \xdef#1{\the\secno.\the\meqno}%
     \global\advance\meqno by1
\else\ifnum\appno>0%
  \warnIfChanged#1{\normalizedromappno.\the\meqno}%
  \xdef#1{\normalizedromappno.\the\meqno}%
     \global\advance\meqno by1
\else%
  \warnIfChanged#1{\the\meqno}%
  \xdef#1{\the\meqno}%
     \global\advance\meqno by1
\fi\fi%
\eqlabeL#1%
\ifWritingAuxFile\immediate\write\auxfile{\noexpand\xdef\noexpand#1{#1}}\fi%
}
\def\anoneqn{\makeNormalizedRomappno%
\ifnum\secno>0
  \eqno(\the\secno.\the\meqno)%
     \global\advance\meqno by1
\else\ifnum\appno>0
  \eqno({\rm\normalizedromappno}.\the\meqno)%
     \global\advance\meqno by1
\else
  \eqno(\the\meqno)%
     \global\advance\meqno by1
\fi\fi%
}
\def\mfig#1#2{\ifx#20
\else\global\advance\figno by1%
\relax#1\the\figno%
\warnIfChanged#2{\the\figno}%
\xdef#2{\the\figno}%
\reflabeL#2%
\ifWritingAuxFile\immediate\write\auxfile{\noexpand\xdef\noexpand#2{#2}}\fi\fi%
}

\catcode`@=11 

\newif\ifFiguresInText\FiguresInTexttrue
\newif\if@FigureFileCreated
\newwrite\capfile
\newwrite\figfile

\newif\ifcaption
\captiontrue
\def\captionsize{\tenrm}
\def\PlaceTextFigure#1#2#3#4{%
\vskip 0.5truein%
#3\hfil\epsfbox{#4}\hfil\break%
\ifcaption\hfil\vbox{\captionsize Figure #1. #2}\hfil\fi%
\vskip10pt}
\def\PlaceEndFigure#1#2{%
\epsfxsize=\hsize\epsfbox{#2}\vfill\centerline{Figure #1.}\eject}

\def\LoadFigure#1#2#3#4{%
\ifundef#1{\phantom{\mfig{}#1}}\else
\fi%
\ifFiguresInText
\PlaceTextFigure{#1}{#2}{#3}{#4}%
\else
\if@FigureFileCreated\else%
\immediate\openout\capfile=\jobname.caps%
\immediate\openout\figfile=\jobname.figs%
@FigureFileCreatedtrue\fi%
\immediate\write\capfile{\noexpand\item{Figure \noexpand#1.\ }{#2}\vskip10pt}%
\immediate\write\figfile{\noexpand\PlaceEndFigure\noexpand#1{\noexpand#4}}%
\fi}

\def\listfigs{\ifFiguresInText\else%
\vfill\eject\immediate\closeout\capfile
\immediate\closeout\figfile%
\centerline{{\bf Figures}}\bigskip\frenchspacing%
\catcode`@=11 
\def\captionsize{\tenrm}
\input \jobname.caps\vfill\eject\nonfrenchspacing%
\catcode`\@=\active
\catcode`@=12  
\input\jobname.figs\fi}

\font\ninerm=cmr9
\font\eightrm=cmr8
\font\sixrm=cmr6

\def\loadtrueseventeenpoint{
 \font\seventeenrm=cmr10 at 17.28truept
 \font\seventeeni=cmmi10 at 17.28truept
 \font\seventeenbf=cmbx10 at 17.28truept
 \font\seventeenit=cmti10 at 17.28truept
 \font\seventeensl=cmsl10 at 17.28truept
 \font\seventeensy=cmsy10 at 17.28truept
}
\def\loadfourteenpoint{
\font\fourteenrm=cmr10 at 14.4pt
\font\fourteeni=cmmi10 at 14.4pt
\font\fourteenit=cmti10 at 14.4pt
\font\fourteensl=cmsl10 at 14.4pt
\font\fourteensy=cmsy10 at 14.4pt
\font\fourteenbf=cmbx10 at 14.4pt
}
\def\loadtruetwelvepoint{
\font\twelverm=cmr10 at 12truept
\font\twelvei=cmmi10 at 12truept
\font\twelveit=cmti10 at 12truept
\font\twelvesl=cmsl10 at 12truept
\font\twelvesy=cmsy10 at 12truept
\font\twelvebf=cmbx10 at 12truept
}

\font\ninei=cmmi9
\font\eighti=cmmi8
\font\sixi=cmmi6
\skewchar\ninei='177 \skewchar\eighti='177 \skewchar\sixi='177

\font\ninesy=cmsy9
\font\eightsy=cmsy8
\font\sixsy=cmsy6
\skewchar\ninesy='60 \skewchar\eightsy='60 \skewchar\sixsy='60

\font\ninebf=cmbx9
\font\eightbf=cmbx8
\font\sixbf=cmbx6

\font\ninett=cmtt9
\font\eighttt=cmtt8

\hyphenchar\tentt=-1 
\hyphenchar\ninett=-1
\hyphenchar\eighttt=-1

\font\ninesl=cmsl9
\font\eightsl=cmsl8

\font\nineit=cmti9
\font\eightit=cmti8


\newskip\ttglue
\def\tenpoint{\def\rm{\fam0\tenrm}%
  \textfont0=\tenrm \scriptfont0=\sevenrm \scriptscriptfont0=\fiverm
  \textfont1=\teni \scriptfont1=\seveni \scriptscriptfont1=\fivei
  \textfont2=\tensy \scriptfont2=\sevensy \scriptscriptfont2=\fivesy
  \textfont3=\tenex \scriptfont3=\tenex \scriptscriptfont3=\tenex
  \def\it{\fam\itfam\tenit}\textfont\itfam=\tenit
  \def\sl{\fam\slfam\tensl}\textfont\slfam=\tensl
  \def\bf{\fam\bffam\tenbf}\textfont\bffam=\tenbf \scriptfont\bffam=\sevenbf
  \scriptscriptfont\bffam=\fivebf
  \normalbaselineskip=12pt
  \let\sc=\eightrm
  \let\big=\tenbig
  \setbox\strutbox=\hbox{\vrule height8.5pt depth3.5pt width\z@}%
  \normalbaselines\rm}

\def\twelvepoint{\def\rm{\fam0\twelverm}%
  \textfont0=\twelverm \scriptfont0=\ninerm \scriptscriptfont0=\sevenrm
  \textfont1=\twelvei \scriptfont1=\ninei \scriptscriptfont1=\seveni
  \textfont2=\twelvesy \scriptfont2=\ninesy \scriptscriptfont2=\sevensy
  \textfont3=\tenex \scriptfont3=\tenex \scriptscriptfont3=\tenex
  \def\it{\fam\itfam\twelveit}\textfont\itfam=\twelveit
  \def\sl{\fam\slfam\twelvesl}\textfont\slfam=\twelvesl
  \def\bf{\fam\bffam\twelvebf}\textfont\bffam=\twelvebf%
  \scriptfont\bffam=\ninebf
  \scriptscriptfont\bffam=\sevenbf
  \normalbaselineskip=12pt
  \let\sc=\eightrm
  \let\big=\tenbig
  \setbox\strutbox=\hbox{\vrule height8.5pt depth3.5pt width\z@}%
  \normalbaselines\rm}

\def\fourteenpoint{\def\rm{\fam0\fourteenrm}%
  \textfont0=\fourteenrm \scriptfont0=\tenrm \scriptscriptfont0=\sevenrm
  \textfont1=\fourteeni \scriptfont1=\teni \scriptscriptfont1=\seveni
  \textfont2=\fourteensy \scriptfont2=\tensy \scriptscriptfont2=\sevensy
  \textfont3=\tenex \scriptfont3=\tenex \scriptscriptfont3=\tenex
  \def\it{\fam\itfam\fourteenit}\textfont\itfam=\fourteenit
  \def\sl{\fam\slfam\fourteensl}\textfont\slfam=\fourteensl
  \def\bf{\fam\bffam\fourteenbf}\textfont\bffam=\fourteenbf%
  \scriptfont\bffam=\tenbf
  \scriptscriptfont\bffam=\sevenbf
  \normalbaselineskip=17pt
  \let\sc=\elevenrm
  \let\big=\tenbig
  \setbox\strutbox=\hbox{\vrule height8.5pt depth3.5pt width\z@}%
  \normalbaselines\rm}

\def\seventeenpoint{\def\rm{\fam0\seventeenrm}%
  \textfont0=\seventeenrm \scriptfont0=\fourteenrm \scriptscriptfont0=\tenrm
  \textfont1=\seventeeni \scriptfont1=\fourteeni \scriptscriptfont1=\teni
  \textfont2=\seventeensy \scriptfont2=\fourteensy \scriptscriptfont2=\tensy
  \textfont3=\tenex \scriptfont3=\tenex \scriptscriptfont3=\tenex
  \def\it{\fam\itfam\seventeenit}\textfont\itfam=\seventeenit
  \def\sl{\fam\slfam\seventeensl}\textfont\slfam=\seventeensl
  \def\bf{\fam\bffam\seventeenbf}\textfont\bffam=\seventeenbf%
  \scriptfont\bffam=\fourteenbf
  \scriptscriptfont\bffam=\twelvebf
  \normalbaselineskip=21pt
  \let\sc=\fourteenrm
  \let\big=\tenbig
  \setbox\strutbox=\hbox{\vrule height 12pt depth 6pt width\z@}%
  \normalbaselines\rm}

\def\ninepoint{\def\rm{\fam0\ninerm}%
  \textfont0=\ninerm \scriptfont0=\sixrm \scriptscriptfont0=\fiverm
  \textfont1=\ninei \scriptfont1=\sixi \scriptscriptfont1=\fivei
  \textfont2=\ninesy \scriptfont2=\sixsy \scriptscriptfont2=\fivesy
  \textfont3=\tenex \scriptfont3=\tenex \scriptscriptfont3=\tenex
  \def\it{\fam\itfam\nineit}\textfont\itfam=\nineit
  \def\sl{\fam\slfam\ninesl}\textfont\slfam=\ninesl
  \def\bf{\fam\bffam\ninebf}\textfont\bffam=\ninebf \scriptfont\bffam=\sixbf
  \scriptscriptfont\bffam=\fivebf
  \normalbaselineskip=11pt
  \let\sc=\sevenrm
  \let\big=\ninebig
  \setbox\strutbox=\hbox{\vrule height8pt depth3pt width\z@}%
  \normalbaselines\rm}

\def\eightpoint{\def\rm{\fam0\eightrm}%
  \textfont0=\eightrm \scriptfont0=\sixrm \scriptscriptfont0=\fiverm%
  \textfont1=\eighti \scriptfont1=\sixi \scriptscriptfont1=\fivei%
  \textfont2=\eightsy \scriptfont2=\sixsy \scriptscriptfont2=\fivesy%
  \textfont3=\tenex \scriptfont3=\tenex \scriptscriptfont3=\tenex%
  \def\it{\fam\itfam\eightit}\textfont\itfam=\eightit%
  \def\sl{\fam\slfam\eightsl}\textfont\slfam=\eightsl%
  \def\bf{\fam\bffam\eightbf}\textfont\bffam=\eightbf \scriptfont\bffam=\sixbf%
  \scriptscriptfont\bffam=\fivebf%
  \normalbaselineskip=9pt%
  \let\sc=\sixrm%
  \let\big=\eightbig%
  \setbox\strutbox=\hbox{\vrule height7pt depth2pt width\z@}%
  \normalbaselines\rm}

\def\tenbig#1{{\hbox{$\left#1\vbox to8.5pt{}\right.\n@space$}}}
\def\ninebig#1{{\hbox{$\textfont0=\tenrm\textfont2=\tensy
  \left#1\vbox to7.25pt{}\right.\n@space$}}}
\def\eightbig#1{{\hbox{$\textfont0=\ninerm\textfont2=\ninesy
  \left#1\vbox to6.5pt{}\right.\n@space$}}}

\def\footnote#1{\edef\@sf{\spacefactor\the\spacefactor}#1\@sf
      \insert\footins\bgroup\eightpoint
      \interlinepenalty100 \let\par=\endgraf
        \leftskip=\z@skip \rightskip=\z@skip
        \splittopskip=10pt plus 1pt minus 1pt \floatingpenalty=20000
        \smallskip\item{#1}\bgroup\strut\aftergroup\@foot\let\next}
\skip\footins=12pt plus 2pt minus 4pt 
\dimen\footins=30pc 

\newinsert\margin
\dimen\margin=\maxdimen

\loadtruetwelvepoint 
\loadtrueseventeenpoint

\def\eatOne#1{}
\def\ifundef#1{\expandafter\ifx%
\csname\expandafter\eatOne\string#1\endcsname\relax}
\def\notTrue{\iffalse}\def\isTrue{\iftrue}
\def\ifdef#1{{\ifundef#1%
\aftergroup\notTrue\else\aftergroup\isTrue\fi}}
\def\use#1{\ifundef#1\linemessage{Warning: \string#1 is undefined.}%
{\tt \string#1}\else#1\fi}



%
\catcode`"=11
\let\quote="
\catcode`"=12
\chardef\foo="22
\global\newcount\refno \global\refno=1
\newwrite\rfile
\newlinechar=`\^^J
\def\@ref#1#2{\the\refno\n@ref#1{#2}}
\def\h@ref#1#2#3{\href{#3}{\the\refno}\n@ref#1{#2}}
\def\n@ref#1#2{\xdef#1{\the\refno}%
\ifnum\refno=1\immediate\openout\rfile=\jobname.refs\fi%
\immediate\write\rfile{\noexpand\item{[\noexpand#1]\ }#2.}%
\global\advance\refno by1}
\def\nref{\n@ref} 
\def\ref{\@ref}   
\def\hrref{\h@ref}
\def\lref#1#2{\the\refno\xdef#1{\the\refno}%
\ifnum\refno=1\immediate\openout\rfile=\jobname.refs\fi%
\immediate\write\rfile{\noexpand\item{[\noexpand#1]\ }#2\semi}%
\global\advance\refno by1}
\def\cref#1{\immediate\write\rfile{#1\semi}}

\def\preref#1#2{\gdef#1{\@ref#1{#2}}}

\def\semi{;\hfil\noexpand\break}

\def\listrefs{\vfill\eject\immediate\closeout\rfile
\centerline{{\bf References}}\bigskip\frenchspacing%
\input \jobname.refs\vfill\eject\nonfrenchspacing}

\def\inputAuxIfPresent#1{\immediate\openin1=#1
\ifeof1\message{No file \auxfileName; I'll create one.
}\else\closein1\relax\input\auxfileName\fi%
}
\def\NPB{Nucl.\ Phys.\ B}

\def\PL{Phys.\ Lett.\ }




\newif\ifWritingAuxFile
\newwrite\auxfile
\def\SetUpAuxFile{%
\xdef\auxfileName{\jobname.aux}%
\inputAuxIfPresent{\auxfileName}%
\WritingAuxFiletrue%
\immediate\openout\auxfile=\auxfileName}



\catcode`\@=\active
\catcode`@=12  
\catcode`\"=\active

\def\Box{\mathord{\dalemb{5.9}{6}\hbox{\hskip1pt}}}
\def\dalemb#1#2{{\vbox{\hrule height.#2pt
        \hbox{\vrule width.#2pt height#1pt \kern#1pt \vrule width.#2pt}
        \hrule height.#2pt}}}

\preref\cstwo{  G.\ Chalmers, W.\ Siegel, hep-ph/9708251}

\preref\csone{  G.\ Chalmers, W.\ Siegel, Phys.\ Rev.\ D54:7628, (1996),
hep-th/9606061}

\preref\sdextension{  Z.\ Bern, L.\ Dixon, D.C.\ Dunbar, D.A.\ Kosower,
Phys.\ Lett.\ B394:105
(1997),  hep-th/9611127}

\preref\manpar{  M. Mangano and S.J. Parke, "Phys. Rep.' Š200 (1991) 301
\semi    Z. Bern, L.
Dixon, D.A. Kosower, hep-ph/9602280,
	"Ann. Rev. Nucl. Part. Sci.' Š46 (1996) 109}

\preref\stringbased{  Z. Bern and D. A.\ Kosower, Phys.\ Rev.\ Lett.\
66:1669 (1991) \semi Z.
Bern and D. A.\ Kosower, Nucl.\ Phys 379:451 (1992) \semi Z. Bern and D.
A.\ Kosower, in {\it
Proceedings of the PASCOS-91 Symposium}, eds.\ P. Nath and S. Reucroft
(World Scientific) \semi
Z. Bern and D. C. Dunbar,  Nucl.\ Phys.\ B379:562 (1992)\semi Z. Bern,
UCLA/93/TEP/5,
hep-ph/9304249, proceedings of TASI 1992}

\preref\color{ F. A.\ Berends and W. T.\ Giele, Nucl.\ Phys.\ B294:700
(1987) \semi M.\ Mangano,
Nucl.\ Phys.\ B309:461 (1988)}

\preref\loopcolor{Z. Bern and D. A.\ Kosower, Nucl.\ Phys.\ B362:389 (1991)}

\preref\minireview{ Z.\ Bern, L.\ Dixon, D.A.\ Kosower, Ann.\ Rev.\ Nucl.\
Part.\ Sci.\  46:109
(1996), hep-ph/9602280}

\preref\ashsmol{ A.\ Ashtekar, Phys.\ Rev.\ Lett.\ 57:2244 (1986), Phys.\
Rev. D36:1587 (1987);
T.\ Jacobson and L.\ Smolin, Phys.\ Lett.\ B196:39 (1987), Class.\ Quant.\
Grav.\ 65:583
(1988); J.\ Samuel, Pramana 28:L429 (1987)}

\preref\collinearapp{  Z.\ Bern, G.\ Chalmers, L.\ Dixon, D.A.\ Kosower,
Phys.\ Rev.\ Lett.\
72:2134-2137 (1994), hep-ph/9312333}

\preref\collinearproof{  Z.\ Bern, G.\ Chalmers, Nucl.\ Phys.\ B447:465 (1995),
hep-ph/9503236\semi  G.\ Chalmers, Proceedings of 22nd ITEP Winter School
of Physics, Moscow,
Russia,  22 Feb - 2 Mar 1994, hep-ph/9405393}

\preref\spinorhelicity{  P. De Causmaecker, R. Gastmans, W. Troost, and
T.T. Wu,
	\NPB 206 (1982) 53;
	F.A. Berends, R. Kleiss, P. De Causmaecker, R. Gastmans, W. Troost,
	and T.T. Wu, \NPB 206 (1982) 61;
	Z. Xu, D.-H. Zhang, and L. Chang, \NPB 291 (1987) 392;
	J.F. Gunion and Z. Kunszt, \PL 161 (1985) 333;
	R. Kleiss and W.J. Sterling, \NPB 262 (1985) 235. }

\preref\halpern{ M.\ Halpern, Phys.\ Rev.\ D16: 1798 (1978), Phys.\ Rev.\
D16:3515 (1978),
 Phys.\ Lett.\ 81B:245 (1979), Phys.\ Rev.\ D19:517 (1979)}

\preref\jackiw{ S.\ Deser, R.\ Jackiw, Phys.\ Lett.\ 139B:371 (1984)}

\preref\hertz{ H.\ Hertz, Ann.\ Phys. (Leipzig), 36:1 (1889); O.\ Laporte
and G.E.\ Uhlenbeck Phys.\ Rev.\ 37:1380 (1931)}

\preref\yang{ C.N.\ Yang, Phys.\ Rev.\ 38:1377 (1977);
S.\ Donaldson, Proc.\ Lond.\ Math.\ Soc.\ 50:1 (1985);
V.P.\ Nair and J.\ Schiff, Phys.\ Lett.\ 246B:423 (1990)}

\preref\csprog{  G.\ Chalmers, W.\ Siegel, in progress}


\loadfourteenpoint

\def\half{{1\over 2}}

\baselineskip 17pt

\rightline{ITP-SB-97-75}
\rightline{hep-th yymmdd}

\vskip 2cm

\vglue .5cm
\centerline{\bf T-Dual Formulation of Yang-Mills Theory}

\vskip .3cm

\vglue 1cm
\centerline{{\bf Gordon Chalmers} and {\bf Warren Siegel}}
\vskip .2cm
\baselineskip=13pt
\centerline{\it Institute for Theoretical Physics}
\centerline{\it State University of New York}
\centerline{\it Stony Brook, NY 11794-3840}
\centerline{\it E-mail addresses : chalmers, siegel@insti.physics.sunysb.edu}

\vglue 2cm

\baselineskip=17pt
\centerline{\bf ABSTRACT}
\vglue 0.3cm

We introduce a self-dual field strength which replaces the gauge field in
spontaneously broken Yang-Mills theory, reformulating it as a Lorentz
covariant non-linear sigma model.  This dualized theory is in both a
unitary and
renormalizable gauge:  The self-dual field strength has exactly the three
components  necessary to describe a massive vector field.  In  future work
we shall
utilize this new formulation as a calculational  tool in spontaneously
broken gauge
theories.

\vfill
\break

\loadfourteenpoint

\section{Introduction}

In the recent past there has been much progress in improving the technical
aspects of
perturbative gauge theory calculations.\footnote{${}^\flat$}{These
developments are summarized
in a relatively concise form in [\use\minireview].}   New Feynman rules
derived from  string
theory (i.e. Bern-Kosower rules) [\use\stringbased], unitarity techniques,
analyticity
constraints  on S-matrix elements [\use\collinearapp,\use\collinearproof],
spinor helicity
techniques  [\use\spinorhelicity], and color ordering
[\use\color,\use\loopcolor] have been
indispensible tools in avoiding  much of the algebraic complexity typically
found in the
amplitude calculations.

In a recent work [\use\cstwo ] we furthered this program by introducing a
chiral  formulation of
minimally coupled fermion fields, found by integrating out half of the
components of the  Dirac
spinor in the usual formulation.  The resulting gauge covariantized theory,
in  which only the
self-dual field strength enters explicitly, is

$$  {\cal L}=\psi^\alpha (\Box - m^2) \psi_\alpha +
  \psi^\alpha F_\alpha{}^\beta \psi_\beta \ .
\eqn\fermsimp
$$   The indices $\alpha,\beta$ are Weyl spinor indices, in which only one
chirality
enters in eq.~(\use\fermsimp).  This action leads to a number of technical
advantages in the
calculation of scattering amplitudes involving  external fermions.  Two
noteworthy features of
the new Feynman rules are the  elimination of much of the gamma matrix
algebra directly within
the Lagrangian  in (\use\fermsimp) and the introduction of a reference
momentum for the external
fermion line.  Gauge theory amplitudes with the polarizations of the
external lines closer to
being the same are simpler than usual as can be seen from the spin
independence of the
Lagrangian when truncating to
$F_{\alpha\beta}=0$.   [\use\cstwo].

In this work we perform an analogous manipulation for the gauge field.
Pure  Yang-Mills
theory may be written using a symmetric (anti-) self-dual field
$G_{\alpha\beta}$ as
[\use\ashsmol]

$${\cal L} = {\rm Tr}~ \left[
 {1\over 2} G^{\alpha\beta} G_{\alpha\beta} +
   G^{\alpha\beta} F_{\alpha\beta} \right] \ ,
\eqn\gftheory
$$   where upon functionally integrating out $G_{\alpha\beta}$ we obtain
the usual
formulation of Yang-Mills theory.  However, we may alternatively integrate
out the gauge
field
$A_{\alpha\dot\beta}$ in the presence of a  Higgs effect, and the new
Lagrangian may be used
as a starting point to find further  simplifications in gauge theory
calculations.  For
example, the propagator in this case will be shown in a simultaneously unitary
(no ghosts) and renormalizable  (i.e. $1/(p^2+m^2)$) gauge as
$G_{\alpha\beta}$ has precisely the number of  components to desccribe a
massive vector
field (i.e. $3$).  (This distinguishes it from earlier  field strength
formulations that
used the full non-self-dual field strength [\use\halpern]; further work
along these
lines has also previously been performed in three dimensions
[\use\jackiw].)   The
action in (\use\gftheory), when dropping the ${\cal L} = {1\over 2} {\rm
Tr}~\int d^4x~
G^{\alpha\beta} G_{\alpha\beta}$ component of the Lagrangian, gives rise to
a self-dual
field theory  perturbatively solved in [\use\csone].

We outline this work as follows.  In section 2 we illustrate our technique
first
with the Stueckleberg model followed by the abelian Higgs model.  In
section 3 we
examine the  dualization of the non-abelian theory with couplings to
matter.  We close in
section 4 with a discussion of related and future work.

\section{Abelian Dualization}
\vskip .2in

In this section we take as a simplest case the Abelian vector theory and
demonstrate the
dualization by exchanging the vector  field $A_{\alpha\dot\alpha}$ with a(n)
(anti-) self-dual field strength  $G_{\alpha\beta}$ (which is symmetric
under the interchange
of indices).  The
modified action becomes non-polynomial in the matter fields.  The
specification of the external line
factors for the dualized vector field $G_{\alpha\beta}$ is singular in the
massless limit, but leads to
similar simplifications as one obtains through the use of the reference
momenta in
[\use\spinorhelicity].

\vskip .2in
\subsection{Stueckelberg Theory}

The simplest example to consider is that of the Stueckleberg theory,
$$
{\cal L} = {1\over 2} G^{\alpha\beta}G_{\alpha\beta} +
G^{\alpha\beta}F_{\alpha\beta}
 + {m^2\over 2} A^{\alpha\dot\alpha} A_{\alpha\dot\alpha} \ ,
\eqn\stulag
$$
where the self-dual field strength is defined by $F_{\alpha\beta}=\half
\partial_{{\dot\gamma}(\alpha} A_{\beta)}{}^{\dot\gamma}$.  Integrating
out the field $G^{\alpha\beta}$ gives the Stueckelberg model.  In the
following we
will examine this model and in the next section the abelian Higgs system.

We dualize the vector potential by first varying the Lagrangian in
(\use\stulag)
with respect to the gauge field,
$$
{\delta{\cal L}\over \delta A^{\alpha\dot\alpha}} =0 \quad\rightarrow\quad
 A^{\alpha\dot\alpha} = {1\over m^2} \partial_\gamma{}^{\dot\alpha}
G^{\gamma\alpha} \ .
\eqn\stufieeqn
$$
The field strength $G_{\alpha\beta}$ appears here in the same way as the
Hertz potential
[\use\hertz] except for the fact that it is self-dual.  We use
eq.~(\use\stufieeqn) to eliminate the gauge field
$A^{\alpha\dot\alpha}$  appearing in eq.~(\use\stulag).  Performing this
gives our dualized theory,
$$
{\cal L} = {1\over 2 m^2} G^{\alpha\beta} \left( \Box - m^2 \right)
G_{\alpha\beta} \ .
\eqn\dualstu
$$
The Lagrangian for the massive vector field in eq.~(\use\dualstu) is in a
renormalizable and ghost-free gauge.

The same theory coupled as well to fermions through

$$
{\cal L}=i{\bar\psi}^{\dot\alpha} \nabla_{\dot\alpha\alpha}\psi^\alpha
+ m_\psi\left( \psi^\alpha\psi_\alpha +
{\bar\psi}^{\dot\alpha}{\bar\psi}_{\dot\alpha}\right) \ ,
$$
i.e., massive QED, gives us the field equation
$$
-\partial_\gamma{}^{\dot\alpha} G^{\gamma\alpha} + m^2 A^{\alpha\dot\alpha} +
 i e{\bar\psi}_{\dot\alpha} \psi_\alpha =0 \ ,
$$
and dualized theory,
$$
\eqalign{
{\cal L} = {1\over 2} & G^{\alpha\beta} \left( \Box - m^2 \right)
G_{\alpha\beta}
+ \bar\psi^{\dot\alpha} i\partial_{\alpha\dot\alpha} \psi^\alpha +
 m_\psi\left( \psi^\alpha\psi_\alpha +
{\bar\psi}^{\dot\alpha}{\bar\psi}_{\dot\alpha}\right)
\cr &
 + i{e\over m}
 \bar\psi^{\dot\alpha} \bigl( \partial_{\gamma\dot\alpha}
G^\gamma{}_\alpha\bigr) \psi^\alpha
+ \left({e\over m}\right)^2 {\bar\psi}^{\dot\alpha} {\bar\psi}_{\dot\alpha}
\psi^\alpha \psi_\alpha \ ,
}
\eqn\fermions
$$
where we have rescaled the mass parameter out of the $G$ field through
$G^{\alpha\beta}\rightarrow
mG^{\alpha\beta}$.  The covariant derivatives are defined by
$\nabla_{\alpha\dot\alpha} =
\partial_{\alpha\dot\alpha} - ie A_{\alpha\dot\alpha}$.   We may
alternatively use the abelian Higgs model
to derive the couplings to fermions,  which will be presented in the next
section.  Furthermore, it is
interesting that in the dualized theory the coupling constant appears only
in the
combination $e/m$.

\vskip .2in
\subsection{Abelian Higgs Model}
\vskip .2in

In this section we examine massive vector fields through the Higgs mechanism.
The general (first-order) form of the gauge theory Lagrangian
possessing spontaneous symmetry breaking is

$$  {\cal L}= -\half G^{\alpha\beta}G_{\alpha\beta} + i G^{\alpha\beta}F_{\alpha\beta} 
 + {\bar\psi}^{\dot\alpha} i \nabla_{\alpha\dot\alpha}
\psi^\alpha
 + \bigl( \nabla_{\alpha\dot\alpha} \phi \bigr)^\ast
   \bigl( \nabla^{\alpha\dot\alpha} \phi \bigr)  +  {\lambda\over 4!}
(\vert\phi\vert^2-v^2)^2
\ ,
\eqn\fundlag
$$
We have introduced the complex scalar together with a potential used to
generate
spontaneous symmetry breaking.   Also, we have rescaled the field
$G_{\alpha\beta}$
by a factor of $i$ to eliminate extraneous factors of $i$ from appearing in
the Feynman
rules.  We  use the first order field equation for $A_{\alpha\dot\alpha}$,
i.e.
${\delta {\cal L}/ \delta A^{\alpha\dot\alpha}} = 0$, to  eliminate the
gauge field:

$$ ~ -i\partial_\gamma{}^{\dot\alpha} G^{\alpha\gamma} - e i \phi^\ast
 \nabla^{\alpha\dot\alpha} \phi   +  e i\phi \nabla^{\alpha\dot\alpha,\ast}
\phi^\ast
 - e{\bar\psi}^{\dot\alpha} \psi^\alpha = 0
\eqn\veceqn
$$ Using eq.~(\use\veceqn) we may express the Lagrangian in
eq.~(\use\fundlag)  in terms of the
fields $G_{\alpha\beta}$ (as well as the matter  fields $\phi$ and $\psi$).
We
first choose the gauge where $\phi=\phi^\ast$.  We find upon
substituting the field $A_{\alpha\dot\alpha}$ into the components of the
Lagrangian the terms,

$$    {\cal L}= -\half G^{\alpha\beta} G_{\alpha\beta}
+ \partial_{\alpha\dot\alpha}\phi \partial^{\alpha\dot\alpha}\phi+
 {\bar\psi}^{\dot\alpha}i\partial_{\alpha\dot\alpha} \psi^\alpha
 - {1\over \phi^2} V^{\beta\dot\gamma} V_{\beta\dot\gamma} \ ,
\eqn\dualvec
$$
where
$$
V^{\beta\dot\gamma} = {1\over e}i \partial_\alpha{}^{\dot\gamma}
G^{\alpha\beta}
 + {\bar\psi}^{\dot\gamma} \psi^\beta \ .
$$
We last expand around the true vacuum, $\phi={\tilde\phi}+v$ in order to
generate
the Feynman rules for the broken gauge theory.  The dualized theory in
eq.(\use\dualvec) derived from the original
$G_{\alpha\beta}F^{\alpha\beta}$  coupling is suitable for a Taylor
expansion around the vacuum
$\langle\tilde\phi\rangle=0$.

The unbroken abelian theory is naively recovered
from the massive  one after taking $v=0$.  The Lagrangian is,
after dropping the potential for the scalar field,

$$   {\cal L}= -\half G^{\alpha\beta}G_{\alpha\beta} +
iG^{\alpha\beta}F_{\alpha\beta}
 + {\bar\psi}^{\dot\alpha} i\nabla_{\alpha\dot\alpha} \psi^\alpha
 + (\nabla_{\alpha\dot\alpha} \phi)^\ast (\nabla^{\alpha\dot\alpha} \phi) \ .
\eqn\abelian
$$
We specify the gauge $\phi=\phi^\ast$; the dualized Lagrangian is then

$$   {\cal L}'= \partial_{\alpha\dot\alpha} \phi
\partial^{\alpha\dot\alpha} \phi
- \half G^{\alpha\beta} G_{\alpha\beta} + {\bar\psi}^{\dot\alpha}
 i\partial_{\alpha\dot\alpha} \psi^\alpha
 - {1\over 2\phi^2} \Bigl[ {1\over e}i\partial_\beta{}^{\dot\alpha}
G^{\alpha\beta}
 + {\bar\psi}^{\dot\alpha} \psi^\alpha \Bigr]^2 \ .
\eqn\abdual
$$  Although one may dualize the vector potential in this manner, the
theory in
eq.(\use\abdual) does not admit a perturbative expansion  because the
scalar field enters
through the denominator in the last  term, and the action is both
nonpolynomial
and singular.  In the previous example of a
spontaneously broken gauge theory the singularity in $\phi$ is cured.  In the
spontaneously broken theory, the three components of the symmetric field
$G_{\alpha\beta}$  naturally describe the degrees of freedom of a massive
vector field.

We may formally define
the scattering amplitudes in the unbroken theory by taking  the vacuum
value $\langle
\phi\rangle$ to zero in the amplitudes derived  from eq.~(\use\abdual).  Of
course, one must be
careful in defining this  limit because one component of the three states
modeled by
$G_{\alpha\beta}$  decouples in the massless limit.  The actual amplitude
calculations  derived
from the couplings in  (\use\dualvec) for the massive vector field should
limit appropriately to the massless case upon taking the scalar vacuum
value to zero.   The
free field equation for the $G_{\alpha\beta}$ is

$$   \left[ \Box-\left( ev\right)^2 \right]
  G_{\alpha\beta}=0 \ ,
\eqn\freefield
$$    and becomes in the massless limit simply $\Box G_{\alpha\beta}=0$,
i.e.  that for a
massless photon field.  The components of the gauge field  strength
$G_{\alpha\beta}$ describe
in the massive case the three independent  spin components of the massive
vector; in the
massless limit, we may explicitly  solve for the free-field solutions
describing the helicity
states of the  photon: We have from eq.~(\use\veceqn) the zeroth order
equation  for
$A_{\alpha\dot\alpha}$ in terms of $G_{\alpha\beta}$,

$$   A_{\alpha\dot\alpha}={1\over v^2} i\partial^\beta{}_{\dot\alpha}
 G_{\beta\alpha} + {\cal O}(\psi,\phi) \ .
\eqn\agfreerel
$$    Using, for example, the $+$ helicity solution to the gauge field we
may  find the
corresponding free-field state for $G_{\alpha\beta}$ from
eq.~(\use\agfreerel):

$$    A_{\alpha\dot\alpha}^{(+)}(k) = i{q_\alpha k_{\dot\alpha}\over
\langle qk\rangle}
 ~\longrightarrow ~ G_{\alpha\beta}^{(+)}={q_\alpha q_\beta\over \langle
qk\rangle^2} \ .
\anoneqn
$$  The two remaining states are found by orthogonality, e.g.
$G_{\alpha\beta}^{(+)}
G^{(-),\alpha\beta}=-1$, and give for the negative and scalar helicity
components of the
$G_{\alpha\beta}$-field,

$$   G^{(-)}_{\alpha\beta}=-k_\alpha k_\beta  \qquad
G^{(0)}_{\alpha\beta}= {q_{(\alpha}
k_{\beta)} \over \langle qk\rangle} \ .
 \anoneqn
$$   In deriving the massless limit from the massive theory in
eq.~(\use\dualvec)  we should use
the solutions to the massive field equation $\left[\Box-\left(
ev\right)^2\right] G_{\alpha\beta}=0$
that limit to the $\pm$ helicity states described above.

\vskip .2in
\subsection{Feynman Rules - Abelian }

The Feynman rules may be directly found from the Lagrangian
in eqs.~(\use\dualvec) by expanding around
$\tilde\phi=0$.  To order $\tilde\phi^2$ we have

$$   {\cal L}_k= -{1\over  2} G^{\alpha\beta}
   \left[\Box-\left( ev\right)^2\right] G_{\alpha\beta}  +
\partial^{\alpha\dot\alpha}{\tilde\phi}
\partial_{\alpha\dot\alpha}{\tilde\phi} +
{\bar\psi}^{\dot\alpha}i\partial_{\alpha\dot\alpha} \psi^{\alpha}
$$
$$
{\cal L}_i= V(\tilde\phi+v)
 - {1\over 2} \left( \partial_\alpha{}^{\dot\gamma} G^{\alpha\beta} \right)
\left( \partial_{\dot\gamma \mu} G^\mu{}_\beta \right)
  \left[ 2{\tilde\phi\over v}  - \left({\tilde\phi\over v}\right)^2 \right]
$$
$$  + \left[ {1\over v}  \left( i\partial_\alpha{}^{\dot\gamma}
G^{\alpha\beta} \right)
 \left( {\bar\psi}_{\dot\gamma} \psi_\beta \right)
+ {1\over v^2} \psi^\alpha\psi_\alpha
{\bar\psi}^{\dot\alpha}{\bar\psi}_{\dot\alpha} \right]
 \left\{ 1 +
  2{\tilde\phi\over v} - \left({\tilde\phi\over v}\right)^2 \right\} \ .
\eqn\expdual
$$
\medskip\noindent  We have rescaled the field strength
$G^{\alpha\beta}\rightarrow
ev G^{\alpha\beta}$ in obtaining the Lagrangian in eq.~(\use\expdual).  It
may appear
that there are terms of the form
$\partial_\alpha{}^{\dot\gamma}G^{\alpha\beta}
\partial_{\beta\dot\gamma}\phi$  in the Taylor
expansion of eq.~(\use\dualvec); however, these are  trivially zero after
integrating by parts.

The dualized form above in (\use\dualvec) generates around the vacuum all
of the  interactions
obtained from the theory in eq.(\use\fundlag), albeit through an infinite
number of terms
(as in a non-linear sigma model, or super Yang-Mills in superspace).  The
field redefinition in  eq.~(\use\veceqn) has shuffled many interactions
into the definition  of
the $G$-field.  The manipulations we have performed at the level of the
action is similar
to the ones performed in [\use\csone] for the case of fermions; to any
given order
there are fewer contributing terms to a Feynman graph and we expect the
perturbative
calculations of amplitudes to be simpler [\use\csprog].

Several of the Feynman rules will be needed in the following sections;  we
list them in the
following.   The propagator for the $G^{\alpha\beta}$ field gives,
$$
\langle G^{\alpha\beta}(p) G^{\mu\nu}(-p) \rangle =
    {1\over p^2+ ({ve})^2}
 (C^{\alpha\mu} C^{\beta\nu} + C^{\alpha\nu} C^{\beta\mu} )  \ .
\anoneqn
$$ The three-point vertices from the fermion contribution in (\use\dualvec)
and (\use\expdual)  are of the form:

$$
\eqalign{
\langle {\bar\psi}^{\dot\nu}(k_1) \psi_\mu(k_2) G_{\alpha\beta}(k_3) \rangle
  & = {v\over 2} (k_1-k_2)^{\dot\nu}{}_{(\alpha} C_{\beta)\mu}
\cr
\langle \tilde\phi (k_1) G_{\mu\nu} (k_2) G_{\alpha\beta} (k_3) \rangle
  & = {1\over v} k_{2,\alpha}{}^{\dot\gamma} k_{3,{\dot\gamma}\mu}
     \ ,  }
\eqn\threepoint
$$   We may absorb the factor of $v$ into the field
$G_{\alpha\beta}$ giving a form for the couplings more similar to the
ones derived from the Stuckelberg model in (\use\stulag).  The remaining
four-point couplings are easily obtained from the  expanded terms in
eq.~(\use\expdual).

\vskip .2in
\section{Non-Abelian Dualization}

In this section we extend the previous analysis to a non-abelian Yang-Mills
theory coupled to matter.
For simplicity we will in this section consider only the case where  all
vector bosons acquire masses,
hence we will dualize the entire set.  The Lagrangian is
$$
 {\cal L}~ =~  {\rm Tr}~\Bigl( {\half}G^{\alpha\beta}G_{\alpha\beta}
 + G^{\alpha\beta}F_{\alpha\beta} \Bigr)   + {\bar\psi}^{\dot\alpha}
i\nabla_{\alpha\dot\alpha}^{\psi} \psi^\alpha
   + (\nabla_{\alpha\dot\alpha}^{\phi} \phi)^\dagger
    (\nabla^{\phi,\alpha\dot\alpha}\phi)  + V(\phi) \ ,
\eqn\nonabel
$$   where the self-dual field strength is defined by $F_{\alpha\beta}^a=
\partial_{(\alpha}{}^{\dot\gamma} A^a_{\beta)\dot\gamma}
 + igf^{abc} A_{b,(\alpha}{}^{\dot\gamma} A_{c,\beta)\dot\gamma}$ and
$$
\nabla^\psi_{\alpha\dot\alpha}  = \partial_{\alpha\dot\alpha} +
gA_{\alpha\dot\alpha}^{a}~
T_a^\psi~,\qquad
\nabla^\phi_{\alpha\dot\alpha} =
\partial_{\alpha\dot\alpha} + gA_{\alpha\dot\alpha}^{a}~ T_a^\phi\ ,
\eqn\covder
$$   are the representation dependent covariant derivatives.

We consider the dual to the general form of the theory above in
eq.~(\use\nonabel).  The
field equation for $A_{\alpha\dot\alpha}$  gives,

$$
\eqalign{  \partial_\rho{}^{\dot\beta} & G_{a}^{\alpha\rho}
 + ig~ G_{b}^{\alpha\rho} A^{\dot\beta}{}_{\rho,c} f_a{}^{bc}
 + ig {\bar\psi}^{\dot\beta} T_a^\psi\psi^\alpha
\cr &
+ g (T_{a}^{\phi}\phi)^\dagger \bigl(\nabla^{\phi,\alpha\dot\beta}
\phi\bigr)   +
g(\nabla^{\phi,\alpha\dot\beta} \phi)^\dagger
  \bigl(T_{a}^{\phi} \phi\bigr)   = 0  \ . }
\eqn\nonabvec
$$
Using the field equation, or alternatively, directly integrating out the gauge
field in the Lagrangian in eq.~(\use\nonabel) gives the dual formulation
$$
{\cal L} = V^{\beta\dot\gamma}_a ~X^{-1}{}^{ab,\alpha}{}_\beta~
V_{b,\alpha\dot\gamma}
+ \partial_{\alpha\dot\alpha}\phi^{a,\ast} \partial^{\alpha\dot\alpha}\phi_a
 + g^2 G^{a,\alpha\beta} G_{a,\alpha\beta}+
 {\bar\psi}^{\dot\alpha,a} \partial_{\alpha\dot\alpha} \psi^\alpha_a   \ .
\eqn\dualnonabellag
$$
The matrix $X$ is defined by
$$
X^{ab}_{\alpha\beta} =  gf^{ab}_c G^c_{\alpha\beta}
 + g^2\phi^\dagger \left\{ T^a,T^b\right\}\phi  ~ C_{\alpha\beta}  \ ,
\eqn\nonabeldual
$$
and the vector $V$ by,
$$
V^{\beta\dot\gamma}_a= \partial_{\mu}{}^{\dot\gamma} G_a^{\beta\mu} +
  g{\bar\psi}^{\dot\gamma} i T_a^\psi \psi^{\beta} +
 g\phi^\dagger T_a^\phi\partial^{\beta\dot\gamma}\phi
$$
with $C^{\alpha\beta}$ and $\eta^{ab}$ used to raise Weyl and group
indices.   We
may power expand the dual theory in eq.~(\use\dualnonabellag) to obtain the
Feynman rules describing the broken phase.  The non-abelian spontaneously
broken
theory will be considered in detail in [\use\csprog], where it will be used
to generate in an efficient manner amplitudes containing massive vectors.

\vskip 1em
\section{Discussion}
\vskip 1em

In this work we have derived from a first-order formulation of a  gauge
theory its
dualized form; the gauge potential is effectively replaced by a self-dual
field strength.
These dual theories are well-defined and admit perturbative  expansions
only in the
case of spontaneous symmetry breaking;  the dual to the unbroken gauge
theory is singular.
The dual formulation expresses Yang-Mills theory as a Lorentz covariant
non-linear sigma model unlike the Yang formulation which expresses the
original gauge field (after gauge fixing) in non-polynomial scalar form for a
self-dual theory [\use\yang].

These dual theories have a number of interesting applications.  First,  the
fact that the dual
to the gauge theory lacks the gauge connection  is interesting for problems
involving
non-trivial solutions that require more  than one patch (i.e. Dirac
monopole).  Second, it is
interesting  that the dual description to Yang-Mills theory may be
understood as  a non-linear
sigma model.  In further work we shall use this  dual formulation as a tool
in doing
perturbative calculations involving  massive vector bosons [\use\csprog].

\vskip 1em
\noindent{\bf Acknowledgements}
\vskip .3in

This work is supported by NSF grant No.~PHY 9722101.

\vfill\break

\listrefs

\end